\documentclass[twoside]{article}
\usepackage{fleqn,espcrc2}

\usepackage{graphicx,graphics}
\usepackage[figuresright]{rotating}

\newcommand{\beq}{\begin{equation}}
\newcommand{\eeq}{\end{equation}}
\newcommand{\bea}{\begin{eqnarray}}
\newcommand{\eea}{\end{eqnarray}}

\newcommand{\e}{\mbox{e}}

\newcommand{\AmS}{{\protect\the\textfont2
  A\kern-.1667em\lower.5ex\hbox{M}\kern-.125emS}}

\title{A Lorentzian cure for Euclidean troubles\thanks{Talk 
presented by R. Loll.}}
\author{J. Ambj\o rn\address{The Niels Bohr Institute, 
Blegdamsvej 17, DK-2100 Copenhagen \O , Denmark\\}\thanks{ 
Supported by ``MaPhySto'', Centre of Mathematical Phy\-sics 
and Stochastics.
}${}^{\S}$,
A. Dasgupta\address{Albert-Einstein-Institut,
Am M\"uhlenberg 1,
D-14476 Golm, Germany},
J. Jurkiewicz\address{Institute of Physics,
Jagellonian University,
Reymonta 4, PL 30-059 Krakow, Poland}${}^{\dag}$\thanks{Supported 
by KBN grant
2P03B 019\,13.}${}^{\S}$
and R. Loll\address{Institute for Theoretical Physics,
Utrecht University, Leuvenlaan 4, NL-3584 CE Utrecht,\\
The Netherlands}\thanks{
Supported by EU network on ``Discrete Random Geometry'', 
grant HPRN-CT-1999-00161, and ESF network no.82 on 
``Geometry and Disorder''.}}
       
\begin{document}

\begin{abstract}
There is strong evidence coming from Lorentzian dynamical
triangulations that the unboundedness of the gravitational
action is no obstacle to the construction of a well-defined
non-perturbative path integral. In a continuum approach, a
similar suppression of the conformal divergence comes about
as the result of a non-trivial path-integral measure.
\vspace{1pc}
\end{abstract}

\maketitle

\section{Lorentzian path integrals for gravity}

The progress of the last few years has established the method of
Lorentzian dynamical triangulations (LDT) as a serious candidate
for a non-perturbative theory of quantum gravity in  four dimensions.
In this approach one tries to define quantum gravity as the
continuum limit of a statistical sum over Lorentzian dynamically
triangulated space-times \cite{al,ajl}. The  transition amplitudes $G$ with
respect to discrete proper time $t$ are given by sums
\beq\label{ampl}
G(\tau_1,\tau_2,t)=\sum_{T,\partial T=\tau_1\cup\tau_2}\frac{1}{C_T}
{\e}^{iS(T)}
\eeq
over inequivalent Lorentzian triangulations $T$ with three-dimensional
spatial boundary triangulations $\tau_1$ and $\tau_2$, weighted by
the gravitational Einstein action of $T$ in Regge form, and including a
discrete symmetry factor $C_T$.

Expression (\ref{ampl}) looks unmanageable at first, but can be
converted into a perfectly well-defined real state sum by means of
a non-perturbative Wick rotation which maps each Lorentzian
triangulation uniquely to a Euclidean one,
\beq
T^{\rm lor}\mapsto T^{\rm eu},
\eeq
and the associated amplitudes to real Boltzmann weights
according to
\beq
{\e}^{i S^{\rm lor}(T^{\rm lor})}\mapsto
{\e}^{- S^{\rm eu}(T^{\rm eu})}.
\eeq
What is more, the Wick-rotated version of the sum (\ref{ampl}) over 
infinitely many Lorentzian triangulated manifolds {\it converges} for 
sufficiently large (positive) bare cosmological constant.

It has already been shown in space-time dimensions two and three 
that the Lorentzian approach is inequivalent to the older method
of Euclidean dynamical triangulations, whose starting point is
a set of Euclidean and not of Lorentzian space-times. Last year's
plenary talk at Lattice 2000 on the subject \cite{plen} contained a
more detailed account of why in higher dimensions the Lorentzian 
approach seems to be preferable.

\section{The conformal-factor problem}

Any non-perturbative path-integral approach to gravity using complex
weights $\e^{iS}$ {\it must} answer the question of how it achieves
convergence, and any path integral using Boltzmann weights $\e^{-S}$
in dimension $d\geq 3$ {\it must} address the question of how it deals
with the unboundedness of the gravitational action. 

The latter problem arises through the unusual behaviour of the 
conformal mode $\lambda(x)$ which 
determines the local scale factor of the metric $g_{\mu\nu}(x)$. 
Isolating its contribution to the kinetic part of the gravitational action,
one finds that it appears with the {\it wrong} sign, rendering the
action unbounded from below. This is most easily seen by applying
a conformal transformation ({\it not} usually a gauge
transformation) to the Einstein action 
\beq
S=k\int d^dx \sqrt{g} (R+\ldots)
\eeq
in dimension $d\geq 3$. Under $g_{\mu\nu}\rightarrow 
g_{\mu\nu}' =\e^\lambda g_{\mu\nu}$, one may write the resulting
new action as
\beq
S'=\int d^dx \sqrt{g'} (-(\partial_0\lambda)^2 +\ldots ).
\eeq
The corresponding Euclidean weight factor is therefore of 
the form
\beq
\e^{-S'}=\e^{\int (\dot\lambda^2 +\ldots )},
\eeq
which can grow without bound, thus 
spelling potential disaster for the Euclidean path integral. 
Perturbatively or in simple minisuperspace models, this
problem is ``fixed'' by adopting some prescription of how the
functional $\lambda$-integration is to be deformed into the
complex $\lambda$-plane. The problem is that such prescriptions
are ad-hoc, non-unique and cannot in any obvious way 
be translated to a non-perturbative context. 

This raises the question of how the non-per\-turbative LDT 
approach -- which after 
Wick-rotating operates with weights $\e^{-S}$ -- overcomes this
difficulty. First, it turns out that as a result of the discretization,
the LDT action is not unbounded below, but has a minimum for
fixed discrete space-time volume $N_d$. For example, in $d=3$ 
the Euclidean action is
\beq
S=N_3\ \Bigl(\frac{k_0}{4}\frac{N_{2,2}}{N_3} +(k_3 -\frac{k_0}{4})\Bigr),
\eeq
and is therefore minimized when the ratio $\tau:= N_{2,2}/N_3$,
$0\leq\tau\leq 1$, of the number of
so-called 2-2 tetrahedral building blocks (c.f. \cite{ajl}) to the total 
number $N_3$ of tetrahedra is minimal. One can construct explicit
triangulations of arbitrary size for which $\tau\simeq 0$. They
correspond to space-times with seemingly large fluctuations of
the conformal factor in proper time, so that neighbouring spatial
slices approximately decouple. 

What must then be investigated {\it dynamically} is whether in the continuum
limit the {\it expectation value} $\langle \tau\rangle$ stays at its
minimum (indicative of a dominance of the kinetic conformal term)
or is bounded away from zero. Our numerical analysis in $d=3$ 
\cite{3d} has shown that $\langle \tau\rangle >0$ for {\it all} finite values
of the inverse gravitational coupling $k_0$! (Note that this is not true
in the Euclidean case where for sufficiently large $k_0$, the path
integral is peaked at configurations with minimal action.) This
means that although ``sick'' configurations with large
conformal excitations are present in the path integral, they are
entropically suppressed and play no role in the continuum limit
of the Lorentzian approach.

It is very interesting to understand whether and how this non-perturbative 
result can possibly be understood from a continuum point of view.
Such an explanation has been provided recently by a continuum
path-integral calculation in proper-time gauge \cite{dl}. Without
attempting to evaluate the full path integral (which is pretty much
impossible because of its non-Gaussian nature), the authors concentrate
on the integral over the conformal factor. This can actually be done 
after borrowing an assumption
on the renormalization of the propagator from a previous successful
application in two-dimensional gravity \cite{dk}. One finds that the
conformal kinetic term in the action is cancelled by a corresponding
contribution from the effective measure, coming from a Faddeev-Popov
determinant which arises during gauge-fixing. Therefore, in agreement with
arguments from a canonical treatment of gravity, the conformal
mode is {\it not} a propagating degree of freedom. As in
the discrete case, it is not just the action contribution, but also the 
non-trivial path integral measure that plays a decisive role in the argument.

In an attempt to study the cancellation mechanism for the conformal
mode explicitly for a Lorentzian dynamically triangulated model, 
various reduced cosmological models in 2+1 dimensions are being
investigated (\cite{dipl} and work in progress). One of the simplest
discrete models one can consider has {\it flat}
tori $T^2$ as its spatial slices at constant integer-$t$. For simplicity,
we can choose to obtain such tori by identifying the opposite
boundaries of a strip of length $l_t$ and width $m_t$ of a regular 
triangulation of 
the flat two-plane so that the associated Teichm\"uller parameter 
$\tau_t$ is purely imaginary ($\tau_t =i m_t/l_t$). 

The Einstein action associated
with a ``space-time sandwich'' $[t,t+1]$ is given by
\bea
S[t,t+1]=
\alpha (l_t +l_{t+1})(m_t+m_{t+1}) 
\nonumber\\
-\beta (l_t-l_{t+1})(m_t-m_{t+1}),
\label{cosm}
\eea
where the (positive) couplings $\alpha$ and $\beta$ are functions of
the bare gravitational and cosmological coupling constants. 
We see that even this simple model suffers potentially from a
conformal factor problem, due to the presence of the
difference term in (\ref{cosm}), leading to a weight factor
\beq
\e^{-S}\sim \e^{\beta \dot l\dot m}.
\eeq
Clearly this term is maximized by having spatial slices
of minimal length and width alternating with slices of maximal
$l$ and $m$. In line with our previous argument we expect that a
cancellation of this conformal divergence can only be achieved
if we allow for sufficiently many interpolating 3d geometries in
between the slices of integer-$t$, so that there is a chance of
``entropy winning over energy''. One can show that for the most
restrictive way of interpolating (requiring that also the sections
at {\it half}-integer times should be flat) this does not happen.
There is not enough entropy and the model does not possess
a continuum limit.

\section{Summary} 

Any gravitational path integral that uses Euclidean weight factors
must address the question of how it deals with the potential
problems caused by the unboundedness of the gravitational action.
Since this situation arises in Lorentzian
dynamical triangulations after one has performed the Wick rotation,
it is of great interest to understand the role of the conformal
divergence in this context.

We have presented strong evidence from both discrete and
continuum approaches that in a proper, non-perturbative 
formulation the conformal sickness seen in perturbative or symmetry-reduced
treatments of the gravitational path integral is absent.
This can happen because the kinetic conformal term in the
action is compensated by non-trivial contributions coming from the
path-integral measure. 

This rather satisfactory result shows that there is nothing wrong in
principle with path-integral formulations of quantum gravity. It is
in line with expectations from canonical formulations where at
least at the classical level the conformal mode is unphysical and
non-propagating. It also suggests that problems in making minisuperspace 
path integrals well-defined may have their root in the ``lack of
entropy" as a result of imposing too stringent symmetry assumptions.

In the case of dynamical triangulations, starting from discrete
geometries of {\it Lorentzian} signature seems to be crucial for
achieving a cancellation of the conformal divergence. This result
gives us further confidence in the method of Lorentzian
dynamical triangulations as a pathway to a theory of
four-dimensional quantum gravity.


\begin{thebibliography}{9}
\bibitem{al} J.\ Ambj\o rn and R.\ Loll, Nucl.\ Phys.\ B536 (1998) 407 
[hep-th/9805108].
\bibitem{ajl} J.\ Ambj\o rn, J.\ Jurkiewicz and R.\ Loll, 
Phys.\ Rev.\ Lett.\ 85 (2000) 924 [hep-th/0002050]; 
Nucl.\ Phys.\ B610 (2001) 347 [hep-th/0105267].
\bibitem{plen} R.\ Loll, Nucl.\ Phys.\ B\ (Proc.\ Suppl.)\ 94 (2001) 96 
[hep-th/0011194].
\bibitem{3d} J.\ Ambj\o rn, J.\ Jurkiewicz and R.\ Loll, 
Phys.\ Rev.\ D64 (2001) 044011 [hep-th/0011276]; JHEP 09 (2001) 022 
[hep-th/0106082].
\bibitem{dl} A.\ Dasgupta and R.\ Loll, Nucl.\ Phys.\ B606 (2001) 357 
[hep-th/0103186].
\bibitem{dk} J.\ Distler and H.\ Kawai, Nucl.\ Phys.\ B32 (1989) 509.
\bibitem{dipl} C. Dehne, Diploma Thesis, University of Hamburg, June 2001.
\end{thebibliography}
\end{document}